\documentclass{appolb}
\usepackage{graphicx}
\usepackage [cp1250]{inputenc}
\usepackage{bm}
\usepackage{bbold}
\usepackage{amsmath}
\usepackage{amssymb}
\usepackage{mathrsfs}

\usepackage{graphicx}

\usepackage{lipsum}    
\usepackage{textcomp}
\usepackage[T1]{fontenc}

\usepackage{dcolumn}
\usepackage{multirow}

\usepackage[usenames, dvipsnames]{color}

\usepackage{url}
\usepackage[colorlinks=true,
pagebackref=true,
linkcolor=blue,anchorcolor=blue,citecolor=blue,filecolor=blue,menucolor=blue,runcolor=blue,
urlcolor=VioletRed
]{hyperref}

\newcommand{\bderr}[2]{\left(\frac{d{#1}}{d{#2}}\right)}

\newcommand{\inbderr}[2]{\left(d{#1}/d{#2}\right)}

\newcommand{\txt}[1]{\textrm{#1}}

\begin{document}
\title{Measuring the speed of sound \\using cumulants of baryon number%
\thanks{Presented at the XXIXth International Conference on Ultra-relativistic Nucleus-Nucleus Collisions, April 2022}%
}
\author{Agnieszka Sorensen, Dmytro Oliinychenko, Larry McLerran
\address{Institute for Nuclear Theory, University of Washington, Box 351550, Seattle, Washington 98195, USA}
\\[3mm]
{Volker Koch 
\address{Lawrence Berkeley National Laboratory, 1 Cyclotron Road, Berkeley, California 94720, USA}
}
}
\maketitle
\begin{abstract}
We show that the values of the first three cumulants of the baryon number distribution can be used to calculate the isothermal speed of sound and its logarithmic derivative with respect to the baryon number density. We discuss applications of this result to heavy-ion collision experiments and address possible challenges, including effects due to baryon number conservation, differences between proton and baryon cumulants, and the influence of finite number statistics on fluctuation observables in both experiment and hadronic transport simulations. In particular, we investigate the relation between quantities calculated in infinite, continuous matter and observables obtained in simulations using a finite number of particles.
\end{abstract}
  
\section{Introduction}
The speed of sound in dense nuclear matter has been given considerable attention in recent years. Measurements and global analyses of neutron star data \cite{Bedaque:2014sqa,Tews:2018kmu, McLerran:2018hbz, Fujimoto:2019hxv, Annala:2019puf, Essick:2020flb, Miller:2021qha, Gorda:2022jvk,Marczenko:2022jhl} suggest that at moderate and high baryon densities, the speed of sound squared $c_s^2$ may exceed the conformal limit of $1/3$. While more studies dedicated to the cold asymmetric nuclear matter will be necessary to confirm this striking behavior, here we propose a complimentary method of studying the speed of sound in dense (nearly symmetric) nuclear matter using observables accessible in heavy-ion collisions.

We are aware of three constraints on $c_s^2$ obtained with heavy-ion collision data. In \cite{Gardim:2019xjs}, the speed of sound in matter created in ultra-relativistic collisions, where the chemical potential $\mu_B \approx  0$, has been estimated using the proportionality of the entropy density and temperature to the charge particle multiplicity and mean transverse momentum, respectively, yielding $c_s^2 = 0.24 \pm 0.04$, consistent with lattice QCD data \cite{Borsanyi:2013bia}. In \cite{Steinheimer:2012bp}, which used data from collisions probing regions of finite baryon number density $n_B$, requiring the Landau model as well as hybrid hydrodynamics and hadronic transport simulations to reproduce the widths of the negatively charged pion rapidity distribution leads to a prediction for a minimum of $c_s^2$ within the collision energy range $\sqrt{s_{NN}}=4\txt{--}9\ \txt{GeV}$. Finally, in \cite{Oliinychenko:2022uvy}, a Bayesian analysis of flow observables utilizing hadronic transport simulations with flexible vector mean-field potentials (tuned to reproduce chosen values of the incompressibility $K_0$ and $c_s^2$ in given density ranges) constrains $c_s^2$ within $0.47 \pm 0.12$ for $n_B \in (2,3)n_0$ and within $-0.08 \pm 0.14$ for $n_B \in (3,4)n_0$, where $n_0 \approx 0.16 ~ \txt{fm}^{-3}$ is the saturation density of nuclear matter.

In the presented work \cite{Sorensen:2021zme}, we suggest another way of constraining $c_s^2$ based on heavy-ion collision data: by utilizing cumulants of the baryon number distribution. Cumulants are one of the most prominent heavy-ion observables due to their characteristic qualitative behavior in the vicinity of a critical point \cite{Asakawa:2009aj, Stephanov:2011pb}. Below, we will use the fact that also the quantitative behavior of the cumulants follows directly from the underlying EOS and, in particular, may shed light on the isothermal speed of sound squared $c_T^2$.

\section{Connecting cumulants of baryon number and the isothermal speed of sound}

Baryon number cumulants are defined as $\kappa_j=VT^{j-1}\left(d^jP/d\mu_B^j\right)_T$, where $P$ is the pressure and $V$ is the volume. In particular, we have
\begin{eqnarray}
&& \kappa_1  = V n_B ~, \label{cumulant_1} \hspace{8mm} \kappa_2 = VT \bderr{n_B}{\mu_B}_T  \label{cumulants} ~.
\end{eqnarray}
Importantly, cumulants are related to moments of the baryon number distribution, measurable in experiments; e.g., for $j\leq3$, $\kappa_j\equiv\Big\langle\Big(N_B-\big\langle N_B\big\rangle\Big)^j\Big\rangle$. 

The isothermal speed of sound is defined as $c_T^2 \equiv \inbderr{P}{\mathcal{E}}_T$ (where $\mathcal{E}$ is the energy density), and by using Eq.\ \eqref{cumulants} and utilizing Maxwell's relations, it can be written as
\begin{eqnarray}
c_T^2  = \bigg[\frac{T}{\kappa_1}  \bderr{\kappa_1}{T}_{\mu_B} +   \frac{\mu_B}{T} \frac{\kappa_2}{\kappa_1}   \bigg]^{-1} ~,
\end{eqnarray}
directly connecting it to the cumulants. However, estimating the first term in the above equation from experimental data poses a significant challenge. At the same time, the second term becomes dominant whenever $\mu_B /T \gg 1$, so that for matter at high baryon densities we can write
\begin{eqnarray}
c_T^2 \approx \frac{T \kappa_1}{\mu_B \kappa_2} ~.
\label{magic_equation_1}
\end{eqnarray}
In a similar fashion, one can connect cumulants with the logarithmic derivative of $c_T^2$ with respect to $n_B$, 
\begin{eqnarray}
\bigg(\frac{d \ln c_T^2}{d \ln n_B} \bigg)_T + c_T^2
= 1 - \frac{\kappa_3 \kappa_1}{\kappa_2^2}   - c_T^2 \bigg(\frac{d \ln (\kappa_2/T)}{d \ln T}\bigg)_{n_B} ~,
\end{eqnarray}
and neglecting the last term on the right-hand side, which is valid for $\mu_B /T \gg 1$, yields
\begin{eqnarray}
\left(\frac{d \ln c_T^2}{d \ln n_B} \right)_T + c_T^2  \approx 1 - \frac{\kappa_3 \kappa_1}{\kappa_2^2} ~.
\label{magic_equation_2}
\end{eqnarray}

Since formulas in Eqs.\ \eqref{magic_equation_1} and \eqref{magic_equation_2} are only approximate, it makes sense to ask what is the region of validity of these equations. As discussed in \cite{Sorensen:2021zme}, based on comparisons of the exact and approximate expressions for $c_T^2$ and its logarithmic derivative in two models of dense nuclear matter, one can expect Eq.\ \eqref{magic_equation_1} to be valid for small to moderate temperatures, $T \lesssim 100~ \txt{MeV}$, and moderate to high baryon chemical potentials, $\mu_B \gtrsim 600\ \txt{MeV}$, while Eq.\ \eqref{magic_equation_2} is appropriate as long as $\mu_B \gtrsim 200\ \txt{MeV}$. We also note here that in the region of validity of our approximations, the difference between $c_T^2$ and the isentropic speed of sound squared $c_{\sigma}^2$, where the latter is more often discussed in the context of heavy-ion collisions, is not substantial.

\section{Experimental data and discussion}

\begin{figure}[tb]
\centerline{%
\includegraphics[width=0.99\textwidth]{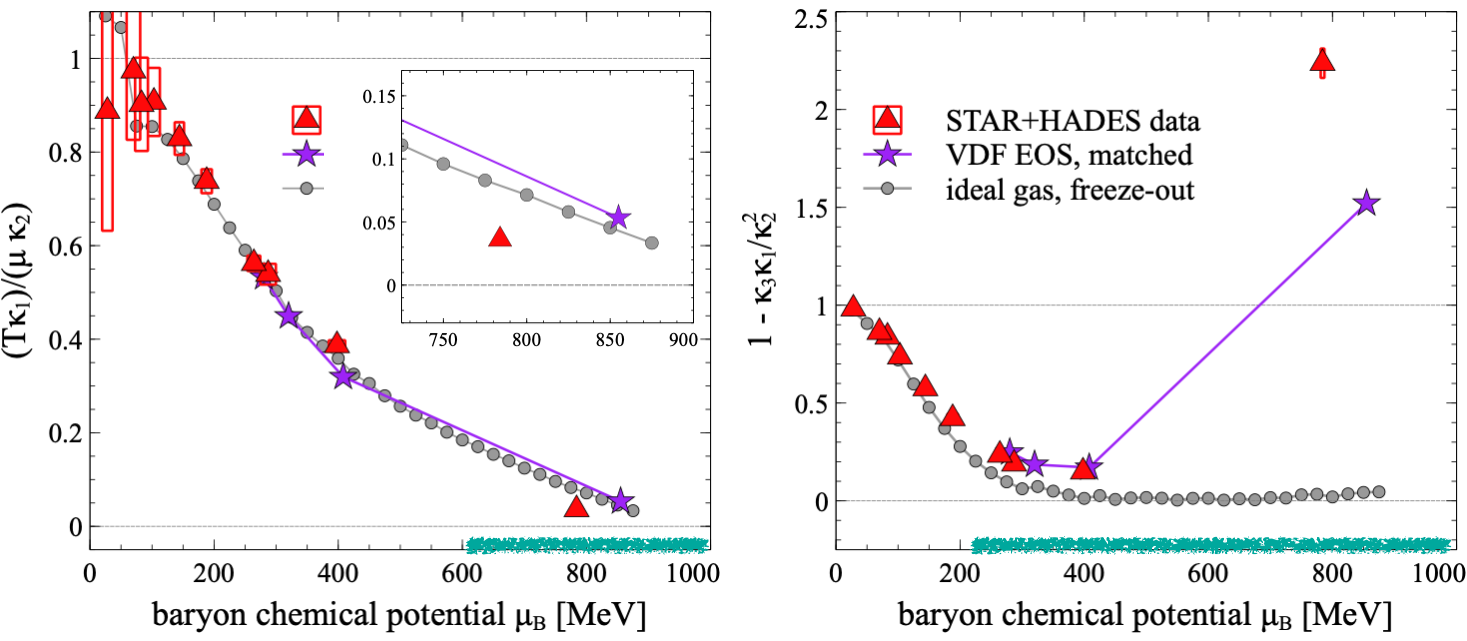} }
\caption{Comparison of the right-hand sides of Eqs.\ \eqref{magic_equation_1} (left panel) and \eqref{magic_equation_2} (right panel) for experimental data (red triangles), the ideal gas at freeze-out (small gray circles), and VDF model results (purple stars); teal lines along the $\mu_B$-axes of the plots indicate the regions of validity of approximations used in Eqs.\ \eqref{magic_equation_1} and \eqref{magic_equation_2}.  
}
\label{Fig:experimental_data}
\end{figure}

To apply our method to heavy-ion collisions, we use experimental data on the cumulants of the net baryon number and chemical freeze-out parameters $(T_{\txt{fo}}, \mu_{\txt{fo}})$, determined by the solenoidal tracker at RHIC (STAR) \cite{Abdallah:2021fzj} and high acceptance dielectron spectrometer (HADES) \cite{Adamczewski-Musch:2020slf, HADESMLorentztalk} in collisions at 0-5\% centrality and for rapidity $|y| < 0.5$ ($|y| < 0.4$ in the case of HADES). \footnote{We note that recently, the STAR collaboration has published results on cumulants of proton number in collisions at $\sqrt{s_{\txt{NN}}} = 3 ~ \txt{GeV}$ \cite{STAR:2021fge}. However, in this measurement a non-symmetric rapidity window of $-0.5 < y <0$ has been used, which does not enable one to make a faithful comparison with the other STAR and HADES data; therefore, we decided against including this data in the current analysis.} With these data, we plot the right-hand sides of Eqs.\ \eqref{magic_equation_1} and \eqref{magic_equation_2} in Fig.\ \ref{Fig:experimental_data} (red triangles); we also show the corresponding results calculated for an ideal gas (small gray circles) and results obtained within the vector density functional (VDF) model \cite{Sorensen:2020ygf}, where the latter have been calculated at points $(T,\mu_B)$ in the phase diagram that allowed us to best reproduce the experimental data. We note here that while the experimental results agree reasonably well with the ideal gas predictions even for small values of $\mu_B$, the fact that these points lie beyond the region of validity means that they do not carry any information about $c_T^2$.

Let us discuss these results. First, as seen in the left panel of Fig.\ \ref{Fig:experimental_data}, the points obtained with the experimental data indicate a relatively small value of the speed of sound, $c_T^2 \approx 0.04$, at large $\mu_B$, $\mu_B \approx 800~\txt{MeV}$ (about half of the corresponding ideal gas value, as shown in the insert). Second, as seen in the right panel of Fig.\ \ref{Fig:experimental_data}, the experimental data indicate a very dramatic change in the behavior of the speed of sound between moderate chemical potentials (shown by experimental results from collisions at $\sqrt{s_{\txt{NN}}} = \big\{14.5, ~ 11.5, ~ 7.7\big\}~\txt{GeV}$) and high chemical potentials (shown by experimental results from collisions at $\sqrt{s_{\txt{NN}}} = 2.4 ~ \txt{GeV}$): while at moderate $\mu_B$ the logarithmic derivative of the speed of sound, $d \ln c_T^2 / d \ln n_B   \approx  1 - \kappa_3 \kappa_1/\kappa_2^2 $, is positive and decreasing, indicating that $c_T^2$ maybe be plateauing as a function of $\mu_B$, at high $\mu_B$, $\mu_B \approx 800~ \txt{MeV}$, we have $1 - \kappa_3 \kappa_1/\kappa_2^2 \approx 2.3$, indicating a very sharp increase in $c_T^2$ with $n_B$ at this region of the phase diagram. As already mentioned in the introduction, such a steep increase in the speed of sound as a function of $n_B$ is consistent with neutron star studies \cite{Bedaque:2014sqa,Tews:2018kmu,Fujimoto:2019hxv, Annala:2019puf, Essick:2020flb, Miller:2021qha, Gorda:2022jvk, Marczenko:2022jhl}, as well as the recent constraint on the speed of sound at $n_B \in (2,3)n_0$ from a Bayesian analysis of flow observables \cite{Oliinychenko:2022uvy}. Interestingly, the matching of the experimental results done within the VDF model (purple stars) suggests that the observed behavior of the cumulants of baryon number and, consequently, the isothermal speed of sound may be primarily due to the ordinary nuclear liquid-gas phase transition instead of the hadron--QGP transition (see Fig.\ 3 in \cite{Sorensen:2021zme} and the associated text); however, more extensive model studies are needed to either confirm or disprove this assertion.

Several comments are in order. First, in our preliminary study we assumed that the cumulants of the net proton number are a good proxy for the cumulants of the net baryon number \cite{Hatta:2003wn}, which is likely too crude of an approximation (see, e.g., \cite{Kitazawa:2011wh}). Furthermore, we assumed that the momentum space cumulants, measured in the experiment, can be used in place of the coordinate space cumulants, Eq.\ \eqref{cumulants}; this is also likely incorrect. Moreover, we did not take the influence of the rapidity bin width into account (which can be done by, e.g., employing finite volume corrections \cite{Vovchenko:2020tsr}). Interestingly, however, a recent, more in-depth analysis \cite{Vovchenko:2022szk} of proton number cumulants measured at $\sqrt{s_{\txt{NN}}} = 2.4 ~ \txt{GeV}$, taking into account some of the issues mentioned above, obtains (using Eqs.\ \eqref{magic_equation_1} and \eqref{magic_equation_2}) $c_T^2 \approx 0.0086  \pm 0.0004$ and $d \ln c_T^2 / d \ln n_B   \approx 5.2 \pm 0.8$, indicating a behavior that qualitatively agrees with our analysis while being even more striking.

Further studies, in particular utilizing hadronic transport simulations, are planned to evaluate to what extent the presented method is applicable to dynamical systems created in heavy-ion collisions and, applying appropriate corrections, to infer the behavior of the speed of sound from cumulants of the baryon number distribution.

\end{document}